\documentclass{article}

\usepackage{arxiv}

\usepackage[utf8]{inputenc} 
\usepackage[T1]{fontenc}    
\usepackage{graphicx, xcolor}
\usepackage{hyperref}
\hypersetup{colorlinks = true, allcolors=.} 
\usepackage{url}            
\usepackage{booktabs}       
\usepackage{lipsum}         
\usepackage{natbib}
\usepackage{doi}

\usepackage{makeidx}         

\usepackage{newtxtext}       %
\usepackage[varvw]{newtxmath}       

\usepackage{caption}
\usepackage{subcaption}

\usepackage{mathtools}
\usepackage{stmaryrd}

\usepackage{amsfonts}       
\usepackage{nicefrac}       
\usepackage{microtype}      
\usepackage{cleveref}       

\usepackage{tikz}
\usetikzlibrary{shapes,arrows}
\usetikzlibrary{intersections}

\usepackage{siunitx}

\title{Simplex space-time meshes in engineering applications with moving domains}

\author{ \href{https://orcid.org/0000-0002-7479-3993}{\includegraphics[scale=0.06]{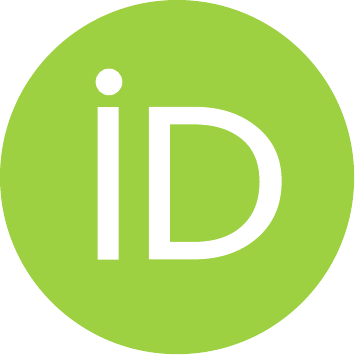}\hspace{1mm}Violeta~Karyofylli}\thanks{\href{https://www.linkedin.com/in/violetakaryofylli/}{\includegraphics[scale=0.06]{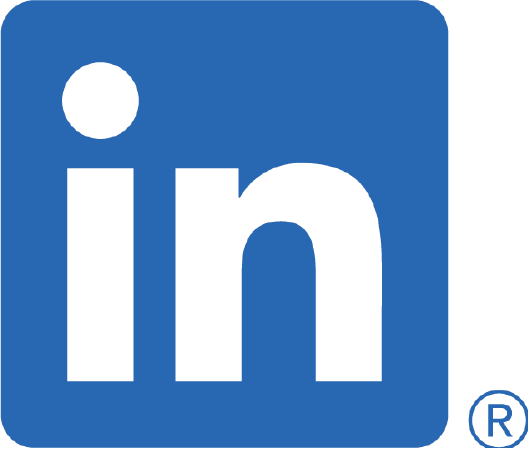}}, \href{https://www.researchgate.net/profile/Violeta-Karyofylli}{\includegraphics[scale=1.0]{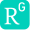}}} \\
	Institute of Energy and Climate Research - \\
	Fundamental Electrochemistry (IEK-9), \\
	Forschungszentrum J{\"u}lich, 52425, Germany \\
	\texttt{v.karyofylli@fz-juelich.de} \\
	\And
	\href{https://orcid.org/0000-0003-4257-8276}{\includegraphics[scale=0.06]{Pictures/orcid.pdf}\hspace{1mm}Marek~Behr} \\
	Chair for Computational Analysis of Technical Systems (CATS), \\
	Center for Simulation and Data Science (JARA-CSD), \\
	RWTH Aachen University, Aachen 52062, Germany \\
	\texttt{behr@cats.rwth-aachen.de} \\
}

\renewcommand{\headeright}{}
\renewcommand{\undertitle}{}
\renewcommand{\shorttitle}{Simplex space-time meshes in engineering applications with moving domains}

\hypersetup{
pdftitle={Simplex space-time meshes in engineering applications with moving domains},
pdfsubject={cs.CE},
pdfauthor={Violeta~Karyofylli, Marek~Behr},
pdfkeywords={space-time, simplex, 4D, finite elements, rotating domains},
}

\begin{document}
\maketitle

\begin{abstract}
This paper highlights how unstructured space-time meshes can be used in production engineering applications with moving domains. Unstructured space-time elements can connect different spatial meshes at the bottom and top level of the space-time domain and deal with complicated domain movements/rotations that the standard arbitrary Lagrangian-Eulerian techniques can not resolve without remeshing. We use a space-time finite element discretization, by means of 4D simplex space-time elements, referred to as pentatopes by \citet{behr2008simplex}, which leads to entirely unstructured grids with varying levels of refinement both in space and in time. Furthermore, we use stabilization techniques, and the stabilization parameter is defined based on the contravariant metric tensor, as shown in the work of \citet{pauli2017stabilized}. Its definition was extended in 4D by \citet{von2019simplex}, allowing us to deal with complex anisotropic simplex meshes in the space-time domain.
\end{abstract}


\section{Introduction}
During the last two decades, the rising interest in 4D simplex space-time (SST) or fully unstructured space-time (UST) discretizations, where the fourth dimension is time, has been apparent. The breakthroughs in computational geometry and, specifically, in the field of triangulations in higher dimensions has also contributed to this growing attention to SST and UST discretizations. For example, based on Whitney's construction, \citet{Boissonnat2020} proposed a robust algorithm for triangulating submanifolds in arbitrary dimensions.

The space-time mesh adaptivity \cite{karyofylli2019simplex,10.1007/978-3-030-70332-5_10,https://doi.org/10.1002/fld.5014}, utilization of parallelism in 4D \cite{neumuller2019time}, and handling of topological changes \cite{neumuller20196} are noteworthy reasons for studying SST or UST discretizations. In the current work, we exploit the advantages of UST discretizations only when dealing with rotating objects inside the space-time domain.

\citet{tezduyar1992newA}, \cite{tezduyar1992newB} were among the first ones who dealt with incompressible flows in deforming domains in the space-time context and developed a robust deforming-spatial-domain/space-time procedure coupled with Galerkin/least-squares. However, they made use of flat space-time (FST) elements for their computations.

Almost two decades later, \citet{neumuller2011refinement} presented an algorithm for arbitrary finite element discretizations of the ST cylinder, resulting in adaptive meshes, movable in time. They computed a pump (2D in space) with a rotating rotor. Moreover, they took advantage of UST discretizations and proposed a refinement scheme in 4D based on the decomposition of a pentatope into smaller pentatopes, which relies on the Freudenthal algorithm \cite{freudenthal1942simplizialzerlegungen}. Almost simultaneously, \citet{rendall2012conservative} extended the space-time approach to the finite volume method and computed unsteady problems allowing any boundary motion or topological change. A fully UST mesh, which can cope with any domain deformation, even with topological changes, was shown by \citet{wang2013discontinuous}. For avoiding re-meshing, they used local mesh operations. \citet{von2019simplex} discretized the compressible Navier-Stokes equations with UST meshes and simulated flow problems on spatial domains that change topology with time. In that work, the compressible flow in a valve that fully closes and opens again was also demonstrated. \citet{von2021four} generated 4D simplex space-time grids with time-variant topology. They based the mesh generation on the extrusion-based approach by \citet{behr2008simplex} and combined it with a four-dimensional extension of the elastic mesh update method (EMUM). \citet{https://doi.org/10.48550/arxiv.2206.01423} constructed four variants of space-time finite element discretizations based on linear tensor-product and simplex-type finite elements and investigated their spatial accuracy and temporal accuracy.

The overall goal of this paper is to provide benchmark cases in two and three space dimensions, dealing with moving domains and simulated with UST discretizations. These results are compared with the ones obtained with an arbitrary Lagrangian-Eulerian (ALE) description, which is naturally included in an FST finite element method. To the author's knowledge, there has not been a similar comparison in four space-time dimensions. In the following sections of this publication, the governing equations, the constitutive law, and the corresponding solution techniques are presented. Additionally, the cases of a rotating stirrer in two and three space dimensions are described and simulated with UST grids. Finally, some concluding remarks are given.

\section{Governing equations}
\label{GoverningEquations}
For the purposes of this paper, we investigate an incompressible, isothermal and single-phase flow and we describe the mathematical model in the following sections.

\subsection{Continuity and momentum equations} \label{Continuity and momentum equations (Two-phase)}
Regarding the mass and momentum conservation, the velocity, \(\mathbf{u}(\mathbf{x}, t )\), and the pressure, \(p(\mathbf{x}, t )\), in the domain \(\Omega_t = \Omega(t)\) need to be computed at every instant \(t \in (t_0 = 0, t_{N}]\). Thus, the continuity and the momentum equations are solved:
\begin{align}
\pmb\nabla\cdot\mathbf{u}&= 0  \label{Continuity_eq} \ \quad\mathrm{and} \\
\rho \left(\frac{\partial \mathbf{u}}{\partial t}+\mathbf{u}\cdot\pmb\nabla\mathbf{u}-\mathbf{f}\right) -\pmb\nabla\cdot\pmb{\sigma}&= \mathbf{0},  \label{Navier_Stokes}
\end{align}
where \(\rho\) is the density, which remains constant. The stress tensor is denoted by \(\pmb{\sigma}\), whereas \(\mathbf{f}\) is the body force per unit mass.

\subsection{Constitutive laws} \label{Constitutive laws (Two-phase)}
The continuity and momentum Equations \eqref{Continuity_eq} -- \eqref{Navier_Stokes} are closed by means of constitutive equations which determine the stress tensor \(\pmb{\sigma}\) as a function of the rate-of-strain tensor, \(\pmb{\varepsilon}(\mathbf{u})\), defined as:
\begin{equation}
\pmb{\varepsilon}(\mathbf{u})= \frac{1}{2}\Big(\pmb\nabla\mathbf{u}+\left(\pmb\nabla\mathbf{u}\right)^T \Big). \label{Strain_tensor}
\end{equation}
The rate-of-strain tensor comes from the symmetric part of the the velocity gradient, \(\pmb\nabla\mathbf{u}\), and expresses the rate of relative movement among neighboring particles. In this paper, we do not use complex constitutive models, but only Newtonian fluids. Thus, we can refer to Equations \eqref{Continuity_eq} and \eqref{Navier_Stokes} as Navier-Stokes equations.

\subsubsection{Newtonian fluids} \label{Newtonian fluids (Two-phase)}
Newtonian models are the simplest constitutive equations. The rate-of-strain tensor \(\pmb{\varepsilon}(\mathbf{u})\), defined in Equation \eqref{Strain_tensor}, is linearly related to a shear stress tensor by a constant effective viscosity \(\mu_{\mathrm{eff}} = \mu \). Combined with the pressure \(p(\mathbf{x}, t )\), the stress tensor \(\pmb{\sigma}\) takes the following form:
\begin{equation}
\pmb{\sigma}(\mathbf{u},p) = -p\mathbf{I}+2\mu \ \pmb{\varepsilon}, \label{Stress_tensor}
\end{equation}
with \(\mathbf{I}\) being the identity tensor.

\subsection{Boundary and initial conditions} \label{Boundary conditions (Two-Phase)}
The essential and natural boundary conditions on \(\Gamma_t = \Gamma(t)\) for the continuity and momentum equations \(\forall \ t \in (t_0 = 0, t_{N}]\) are declared as:
\begin{align}
\mathbf{u}=\mathbf{u}_{g \ } \quad &\mathrm{on} \quad (\Gamma_g)_t  \quad \mathrm{and}  \label{BoundaryUPTConditions1}\\
\mathbf{n} \cdot \pmb{\sigma} ( \mathbf{u}, p)=\mathbf{h} \quad &\mathrm{on} \quad (\Gamma_h)_t, \label{BoundaryUPTConditions2}
\end{align}
where $\mathbf{n}$ is the normal vector on \(\Gamma_t\) and \(\mathbf{u}_g\), and \(\mathbf{h}\) prescribe the velocity and traction, respectively. The Dirichlet and Neumann part of the boundary, \(\Gamma_t\), are represented as \((\Gamma_g)_t\) and \((\Gamma_h)_t\), respectively. They form complementary subsets of \(\Gamma_t\), meaning that \((\Gamma_g)_t \cup (\Gamma_h)_t = \Gamma_t\) and \((\Gamma_g)_t \cap (\Gamma_h)_t = \emptyset\).

Last but not least, the initial conditions for the velocity (divergence-free) have to be defined in \(\Omega_0\) at \(t=0\) for closing the flow problem:
\begin{align}
\mathbf{u}(\mathbf{x},0) &= \mathbf{u}_0(\mathbf{x}). \label{eq:VelocityIC}
\end{align}

\section{Solution technique}
\label{SolutionTechnique}
In order to discretize Equations \eqref{Continuity_eq} and \eqref{Navier_Stokes}, \(\mathbb{P}_1\) simplex space-time (SST) finite elements are used. 
The Galerkin/least-squares (GLS) stabilization method is also applied. In the GLS method, the stabilization term is a least-squares form of the original differential equation, weighted element by element \cite{Donea}. For the creation of the finite element function spaces, which are used for the space-time discretization, the time interval is denoted as \(I = \left(t_0, t_{N}\right]\), with \(t_0 = 0\) and \(t_{N} = T\) being the initial and last time level of the space-time domain. The space-time domain \(Q\) is defined as the region confined by the surfaces \(\Omega_0\) and \(\Omega_{N}\), and represents the evolution of spatial domain over \(I\). A slice of the space-time domain, \(\Omega_{n}\), corresponds to the \(n\)-th time level, \(t_n\), where: \(0 = t_0 < \cdots < t_{n} < \cdots < t_{N} = T\). 
Similarly, the space-time boundary, which is described by the spatial boundary \(\Gamma_0=\Gamma(t_0)\) evolved over \(I\) to its final state, \(\Gamma_N=\Gamma(t_N)\), is referred to as \(P\).
 
The problem is solved in the complete space-time domain, starting with \(\left(\mathbf\bullet^h\right)^+_0=\mathbf\bullet_0\) at \(t_0\), where the superscript \(h\) stands for the discretization and \(\mathbf\bullet\) for the discretized quantities.
In the discretized forms below, the following notation is applied:
\begin{align} 
\left(\mathbf\bullet^h\right)^\pm_0&=\lim_{\varepsilon\to0}\mathbf{\bullet}\left(t_0\pm\varepsilon\right), \label{Notation_1}\\
\int_{Q} \ldots dQ&=\int_{I}\int_{\Omega^h_t} \ldots d\Omega dt \quad \mathrm{and} \label{Notation_2}\\
\int_{P_\mathbf\bullet} \ldots dP&=\int_{I}\int_{\left(\Gamma_\mathbf\bullet\right)^h _t} \ldots d\Gamma dt \label{Notation_3}.
\end{align} 

The finite element interpolation and weighting function spaces which are set for the variables of the equation system (velocity and pressure) are:
\begin{align}
\mathcal{S}_\mathbf{u}^h &= \left\{\mathbf{u}^h \ | \ \mathbf{u}^h \ \in \left[H^{1h}\left(Q\right)\right]^{n_{\mathrm{sd}}}, \ \mathbf{u}^h \ \dot{=} \ \mathbf{u}^h _g \ \mathrm{on} \ P _g \right\}, \label{Interpolation_fs_vel} \\
\mathcal{V}_\mathbf{u}^h &= \left\{\mathbf{w}^h \ | \ \mathbf{w}^h \in \left[H^{1h}\left(Q\right)\right]^{n_{\mathrm{sd}}}, \ \mathbf{w}^h \ \dot{=} \ \mathbf{0} \ \mathrm{on} \ P _g \right\} \ \ \mathrm{and} \label{Weighting_fs_vel} \\
\mathcal{S}_p^h &= \mathcal{V}_p^h = \left\{p^h \ | \ p^h \in L^{2h}\left(Q\right)\right\} , \label{Interpolation_Weighting_fs_pres}
\end{align}
for every space-time slab. Here, \(H^{1h} \subset H^1\) is a finite dimensional Sobolev space consisting of functions which are square-integrable in \(Q\) and have square-integrable first derivatives. The trial function spaces for the velocity, denoted by \(\mathcal{S}_u ^h\), must additionally satisfy the Dirichlet boundary conditions. Similar test function spaces \(\mathcal{V}_u ^h\) are chosen, but it is required that the test functions vanish on the Dirichlet boundary \(P_\mathbf\bullet\). The requirements for the pressure trial function space are less restrictive, as no derivatives of the pressure appear and no explicit pressure boundary conditions exist. Therefore, the pressure trial and test functions are chosen from \(L^{2h} \subset L^2\), being the finite dimensional Hilbert space of square-integrable functions. The interpolation functions in the elements constitute first-order polynomials, which are continuous in space, but discontinuous in time for every region of the domain.

The space-time domain, \(Q\), is discretized into \(n_{\mathrm{el}}\) elements, \(Q^e\), and \(n_{\mathrm{nodes}}\) nodes. The set of all nodes \(\{\mathbf{x}_k\}\) is denoted as \(K\). The shape function associated with the node \(k\) is \(N_k(\mathbf{x},t)\). The \(\mathbb{P}_1\) basis functions are defined by means of the local coordinates, \(\pmb{\xi} = \left\{\xi^1 = \xi, ..., \xi^{n_{\mathrm{sd}} + 1} = \theta\right\}^T\), as
\begin{align}
	N^{(e)}_1(\pmb{\xi})&= 1 - \sum_{d=1}^{n_{\mathrm{sd}}+1}\xi^d, \label{ElementLevelBasisFunctionsOneCh2} \\ 
	N^{(e)}_{d+1}(\pmb{\xi})&= \xi^d, \quad d = 1, ..., n_{\mathrm{sd}} + 1. \label{ElementLevelBasisFunctionsTwoCh2}
\end{align}

Furthermore, in the case of an isoparametric affine linear mapping, \(\Phi\), between the global and local space, which is based on the \(\mathbb{P}_1\) basis functions defined in equations \eqref{ElementLevelBasisFunctionsOneCh2} and \eqref{ElementLevelBasisFunctionsTwoCh2}, the global coordinates, \(\pmb{x} = \left\{x^1 = x, ..., x^{n_{\mathrm{sd}} + 1} = t\right\}^T\), of any interior point of the \(e\)-th element can be expressed as
\begin{equation} \label{IsoparametricElementsCh2}
	\pmb{x} = \Phi(\pmb{\xi}) = \sum_{d=1}^{^{n_{\mathrm{sd}}+2}}N_{d}^{(e)}(\pmb{\xi})\pmb{x}_{d}
\end{equation}
with the \(\pmb{x}_{d}\) being the global coordinates of the \(d\)-th node of an element. Here, it must be pointed out that SST elements have \(n_{\mathrm{sd}}+2\) nodes in total. The computation of the Jacobian of every element, \(\mathbf{J}^e\), can be generalized in \(n_{\mathrm{sd}} + 1\)-dimension as suggested by \citet{von2019simplex, neumuller2011refinement}, and \citet{caplan2019four}:
\begin{equation} \label{JacobianCh2}
	\mathbf{J}^e = \left(\mathbf{x}_2-\mathbf{x}_1, \mathbf{x}_3-\mathbf{x}_1, ..., \mathbf{x}_{n_{\mathrm{sd}}+2}-\mathbf{x}_1\right)
\end{equation}
with the columns of \(\mathbf{J}^e\) being the distance vectors, \(\mathbf{x}_{d+1}-\mathbf{x}_1, \ \forall \ d = 1, ..., n_{\mathrm{sd}} + 1\).

Having already defined the \(\mathbb{P}_1\) basis functions, the velocity and pressure can be approximated as:
\begin{align}
\mathbf{u}^h(\mathbf{x},t) &= \sum_{j=1}^{n_{\mathrm{sd}}} \sum_{k\in K}^{} N_k(\mathbf{x},t)\left(\mathbf{u}_k\cdot\mathbf{e}_j\right)\mathbf{e}_j \label{TrialFunctionsVelocity} \ \ \mathrm{and} \\
p^h(\mathbf{x},t) &= \sum_{k\in K}^{} N_k(\mathbf{x},t)p_k,  \label{TrialFunctionsPressure}
\end{align}
where \(n_{\mathrm{sd}}\) is the spatial dimension and \(e_j\) is a unit vector with a one at the \(j\)-th entry. The nodal values of the velocity and pressure are \(\mathbf{u}_k\) and \(p_k\), respectively. Thus, in the Galerkin formulation, the test functions are defined as:
\begin{align}
\mathbf{w}^h(\mathbf{x},t) =& \sum_{j=1}^{n_{\mathrm{sd}}} w^h _j(\mathbf{x},t)\mathbf{e}_j, \label{TestFunctionsVelocity1} \\
w^h _j(\mathbf{x},t) \in \mathcal{V}_\mathbf{u}^h :=& \mathrm{span}_{k\in K} \left\{N_k(\mathbf{x},t)\right\} \label{TestFunctionsVelocity2} \ \ \mathrm{and} \\
q^h(\mathbf{x},t) \in \mathcal{V}_p^h :=& \mathrm{span}_{k\in K} \left\{N_k(\mathbf{x},t)\right\}. \label{TestFunctionsPressure}
\end{align}

\subsection{Continuity and momentum equations} \label{Solution Technique: Continuity and momentum equations}
The Navier-Stokes Equations \eqref{Continuity_eq} and \eqref{Navier_Stokes} have the following space-time discretized weak formulation after stabilization: 

Given \((\mathbf{u}^h)_0^-\), find \(\mathbf{u}^h \in \mathcal{S}_\mathbf{u}^h\) and \(p^h \in \mathcal{S}_p^h\) such that \(\forall \mathbf{w}^h \in \mathcal{V}_\mathbf{u}^h, \forall q^h \in \mathcal{V}_p^h\):
\begin{align}\label{Variational_Navier-Stokes}
\int_{Q}\mathbf{w}^h&\cdot\rho \left(\frac{\partial \mathbf{u}^h}{\partial t}+\mathbf{u}^h\cdot\pmb{\nabla}\mathbf{u}^h-\mathbf{f} ^h\right) \ dQ + \int_{Q}\pmb{\varepsilon}(\mathbf{w}^h):\pmb{\sigma} (\mathbf{u}^h,p^h) \ dQ \nonumber \\
&+ \int_{Q}q^h\pmb{\nabla}\cdot\mathbf{u}^h\ dQ + \int_{\Omega_0}(\mathbf{w}^h)^+_0\cdot\rho ((\mathbf{u}^h)^+_0-(\mathbf{u}^h)^-_0) \ d\Omega \nonumber \\
&+\sum_{e=1}^{n_{\mathrm{el}}} \int_{Q^e}\tau_{MOM}\frac{1}{\rho } \left[\rho  \left(\frac{\partial \mathbf{w}^h}{\partial t}+\mathbf{u}^h\cdot\pmb{\nabla} \mathbf{w}^h \right)-\pmb{\nabla}\cdot\pmb{\sigma} (\mathbf{w}^h, q^h) \right] \nonumber \\
&\cdot \left[\rho  \left(\frac{\partial \mathbf{u}^h}{\partial t}+\mathbf{u}^h\cdot\pmb{\nabla} \mathbf{u}^h -\mathbf{f} ^h\right)-\pmb{\nabla}\cdot\pmb{\sigma} (\mathbf{u}^h, p^h) \right] \ dQ \nonumber\\
&+\sum_{e=1}^{n_{\mathrm{el}}}\int_{Q^e}\tau_{CONT}\pmb{\nabla}\cdot\mathbf{w}^h\rho \pmb{\nabla}\cdot\mathbf{u}^h\ dQ = \int_{P_h}\mathbf{w}^h\cdot\mathbf{h}^h \ dP. 
\end{align}
The term on the right-hand side of the discretized weak formulation $\int_{P_h}\mathbf{w}^h\cdot\mathbf{h}^h \ dP$, is of primary importance. It can be used for applying traction, \(\mathbf{h}^h\), to a boundary of the domain (as stated by \citet{karyofylli2017novel}). The term \(\mathbf{h}\) corresponds to the imposed traction on the Neumann part of the space-time boundary, $P_h$, as already described in Equation \eqref{BoundaryUPTConditions2}.

\subsection{Stabilization parameters} \label{Solution Technique: Stabilization}
As a final step, the definition for the various stabilization parameters $\tau_{MOM}$ and $\tau_{CONT}$ is needed. The stabilization parameter for the momentum equation is:
\begin{equation}\label{tauMomentum}
\tau_{MOM} = \left(\mathbf{u}^h \cdot \mathbf{G}^{-1} \mathbf{u}^h + C_I \nu^2 \mathbf{G}^{-1}:\mathbf{G}^{-1}\right)^{-\frac{1}{2}},
\end{equation}
where \(\nu= \frac{\mu}{\rho}\) represents the kinematic viscosity of the single-phase fluid. The parameter \(C_I\) is computed based on an inverse estimate inequality and differs among various element types in different dimensions. Moreover, the metric tensor \(\mathbf{G}^{-1}\) is defined as: 
\begin{equation} \label{MetricTensor}
\mathbf{G}^{-1} = \left[\tilde{g}^{ij}\right]=\sum_{k} \frac{\partial \xi^i}{\partial x^k}\frac{\partial \xi^j}{\partial x^k},
\end{equation}
and is considered to be contravariant, following the definition of \citet[Chapter~3]{Karyofylli:816042}. Furthermore, \citet{von2019simplex} proposed a design of this metric tensor, which ensures invariance with respect to node numbering in the case of simplex space-time elements and is also adopted in this work. Initially, this approach was used by \citet{pauli2017stabilized} for flat space-time elements which result from the temporal extrusion of simplicial spatial meshes. We also assume that \(C_I = 1\) for the following simulations. A definition of \(C_I\) can be traced in the work of \citet{knechtges2018simulation}. Regarding the stabilization of the continuity equation, the stabilization parameter is according to \citet{pauli2017stabilized} the following:
\begin{equation}\label{tauContinuity}
\tau_{CONT} = \left(\tau_{MOM}\mathbf{g} \cdot \mathbf{g}\right)^{-1},
\end{equation}
while the components of \(\mathbf{g}\) are declared below:
\begin{equation}\label{tauContinuityNotationOne}
g^i = \sum^{n_{\mathrm{sd}}}_{j=1} \frac{\partial \xi^j}{\partial x^i}.
\end{equation}

\section{Numerical Results}
\label{sec:4}
\begin{figure}[!htb]
	\centering
	\includegraphics[scale=0.50]{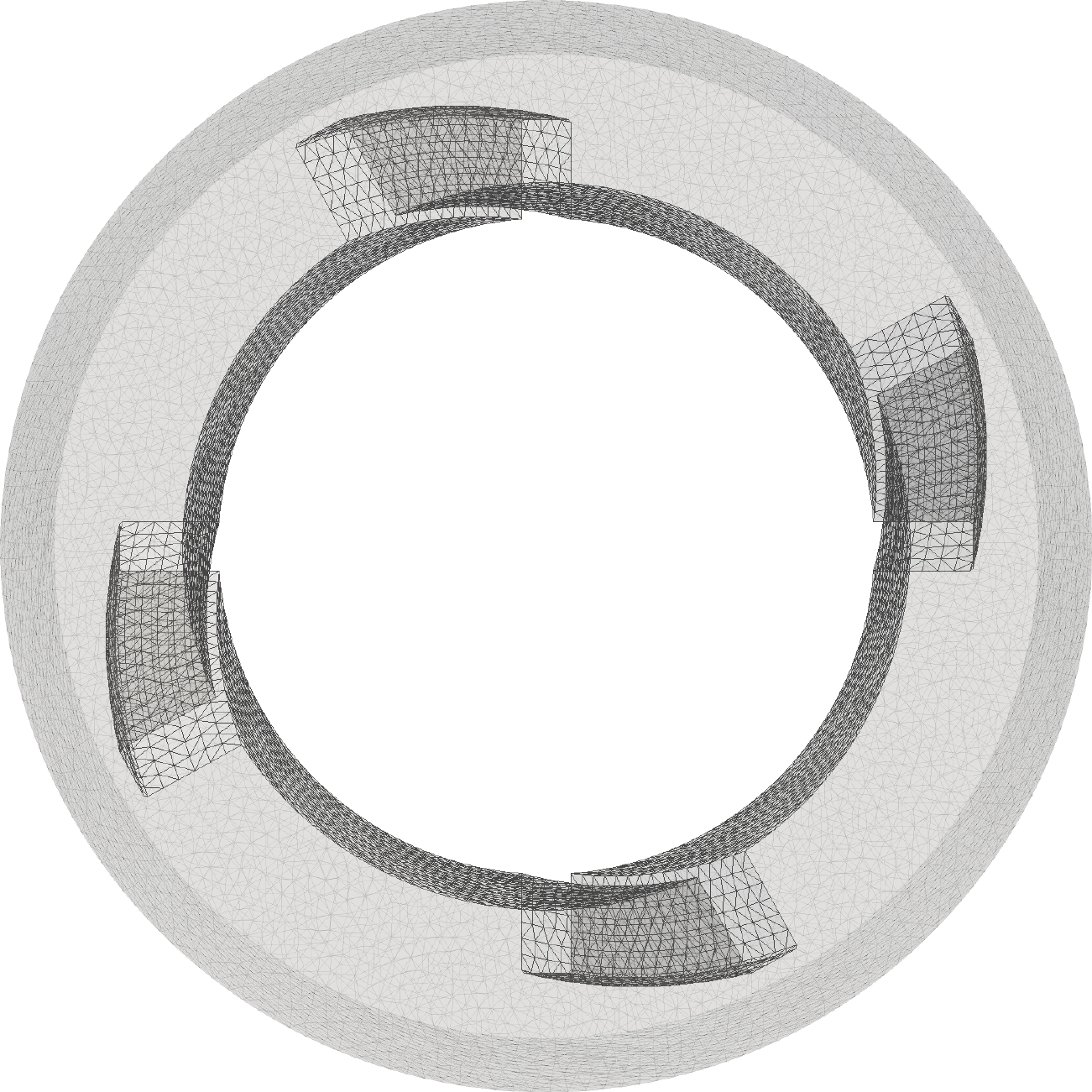}
	\caption{3D space-time mesh of a 2D rottating stirrer.}
	\label{fig:SpaceTimeStirrer3D}
\end{figure}

In this section, we simulate the flow in a circular domain enclosing a rotating stirrer both in two and three space dimensions with the following approaches:
\begin{itemize}
\item a space-time domain that resolves the total movement of the stirrer, as illustrated in Figure \ref{fig:SpaceTimeStirrer3D} for the 2D (in space) case, and
\item the concept of space-time slab in combination with mesh movement techniques.
\end{itemize}
Regarding the former approach, since the motion of the stirrer is pure rotation, we account for it via the construction of the complete space-time domain, $Q$, which is twisted around the time axis. As described in Section \ref{SolutionTechnique}, $Q$ is then decomposed into simplices, and time is treated as an additional dimension in the mesh. Therefore, the movement of the rotor is naturally resolved by the UST discretization. 

Concerning the latter method, its mathematical background was described in detail by \citet{pauli2016stabilized}. The flow field is computed at every possible rotation of the stirrer. For the preservation of a valid mesh during the rotation, the grid around the stirrer is rotated with an equal angular velocity as the stirrer itself, utilizing the ALE description. Furthermore, this approach is based on tensor-product meshes.

\subsection{2D Stirrer}
\label{2DStirrer}

\begin{figure}[!htb]
	\centering
	\begin{tikzpicture}
		\node[anchor=south west,inner sep=0] (image) at (0,0) {\includegraphics[width=0.5\textwidth]{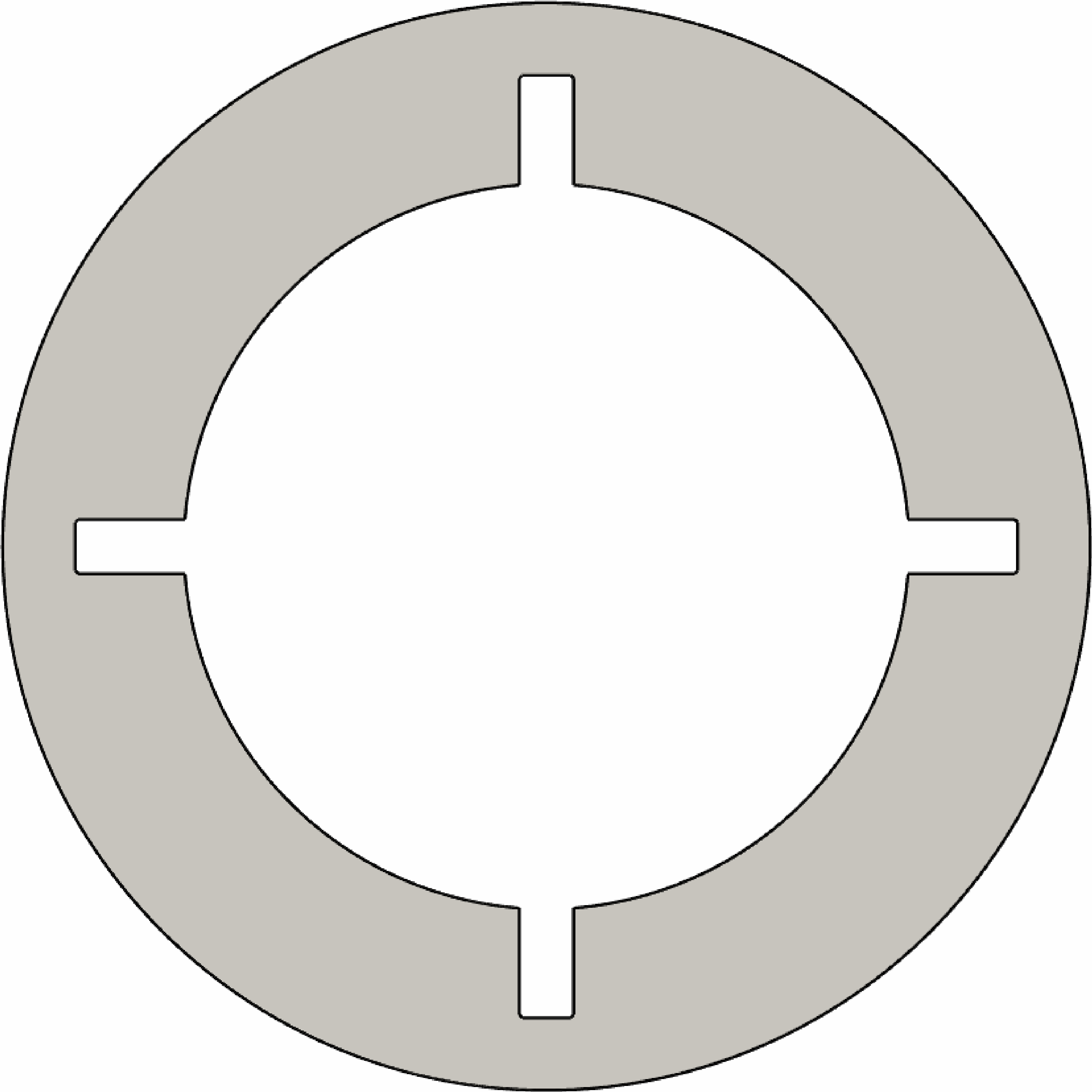}};
		\begin{scope}[x={(image.south east)},y={(image.north west)}]
			\draw[thick, stealth-stealth] (1.05, 0.0) -- node[right] {$\SI{6.0}{}$}  (1.05, 1.0);
			\draw[dashed] (0.5, 1.0) -- (1.05, 1.0);
			\draw[dashed] (0.5, 0.0) -- (1.05, 0.0);
			\draw[thick, stealth-stealth] (0.55, 0.07) -- node[left] {$\SI{5.2}{}$}  (0.55, 0.93);
			\draw[dashed] (0.53, 0.07) -- (0.55, 0.07);
			\draw[dashed] (0.53, 0.93) -- (0.55, 0.93);
			\draw[thick, stealth-stealth] (0.17, -0.05) -- node[below] {$\SI{4.0}{}$}  (0.83, -0.05);
			\draw[dashed] (0.17, -0.05) -- (0.17, 0.5);
			\draw[dashed] (0.83, -0.05) -- (0.83, 0.5);
			\draw[thick, stealth-stealth] (-0.05, 0.47) -- node[left] {$\SI{0.3}{}$} (-0.05, 0.53);
			\draw[dashed] (-0.05, 0.47) -- (0.07, 0.47);
			\draw[dashed] (-0.05, 0.53) -- (0.07, 0.53);
		\end{scope}
	\end{tikzpicture}
	\caption{Dimensions of the 2D stirrer.}
	\label{fig:DimensionsStirrer2D}
\end{figure}
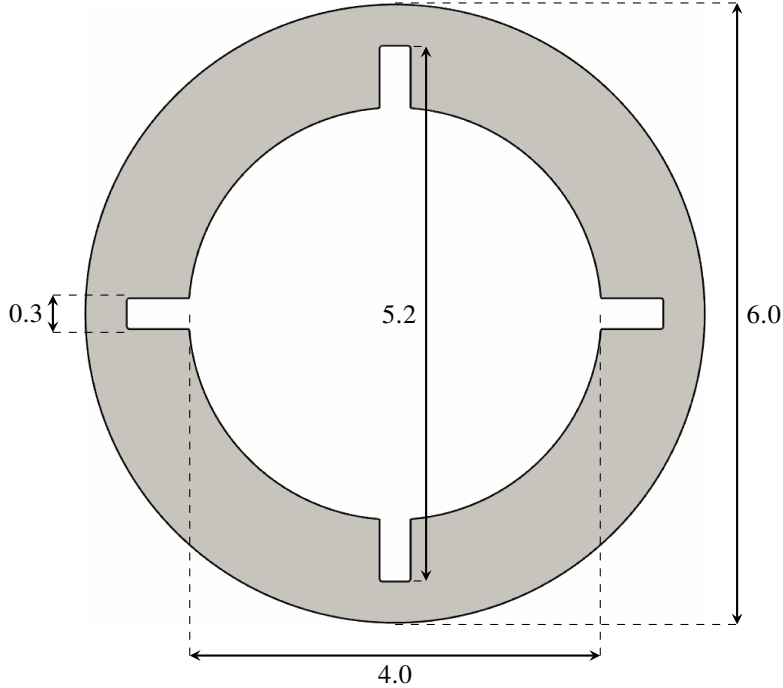
We consider a two-dimensional flow in a circle enclosing a rotating stirrer with four blades as a first test case. The geometry and stirrer dimensions are shown in Figure \ref{fig:DimensionsStirrer2D}. The UST discretization consists of \(\SI{229602}{}\) tetrahedral elements, whereas the spatial grid for the ALE formulation contains \(\SI{4502}{}\) triangles. The stirrer is placed in the center of the circular domain and initially aligned, as shown in Figure \ref{fig:DimensionsStirrer2D}. No-slip boundary conditions are defined on the outer walls of the circle and the stirrer. The stirrer rotates at constant angular velocity counterclockwise. The angular velocity is \(\frac{250}{3}\pi{}\), the density is \(\SI{1.0}{}\), and the dynamic viscosity is \(\SI{0.03382}{}\).

For both computations, the stirrer starts to rotate instantaneously. In the simulation with the UST discretization, we don't need to perform any time marching since time is an additional dimension in the grid, but we perform Newton-Raphson iterations to linearize the Navier-Stokes equations. Here, we need to point out that the system converges within \(\SI{13}{}\) iterations. In the simulation with the concept of space-time slab and ALE formulation, we use a time slab size of \(\Delta t = \SI{0.00012}{}\). One time step usually converges within eight Newton-Raphson iterations.

\begin{figure}[!htb]
	\centering
	\begin{subfigure}{0.475\textwidth}
		\centering
		\includegraphics[width=1.0\linewidth]{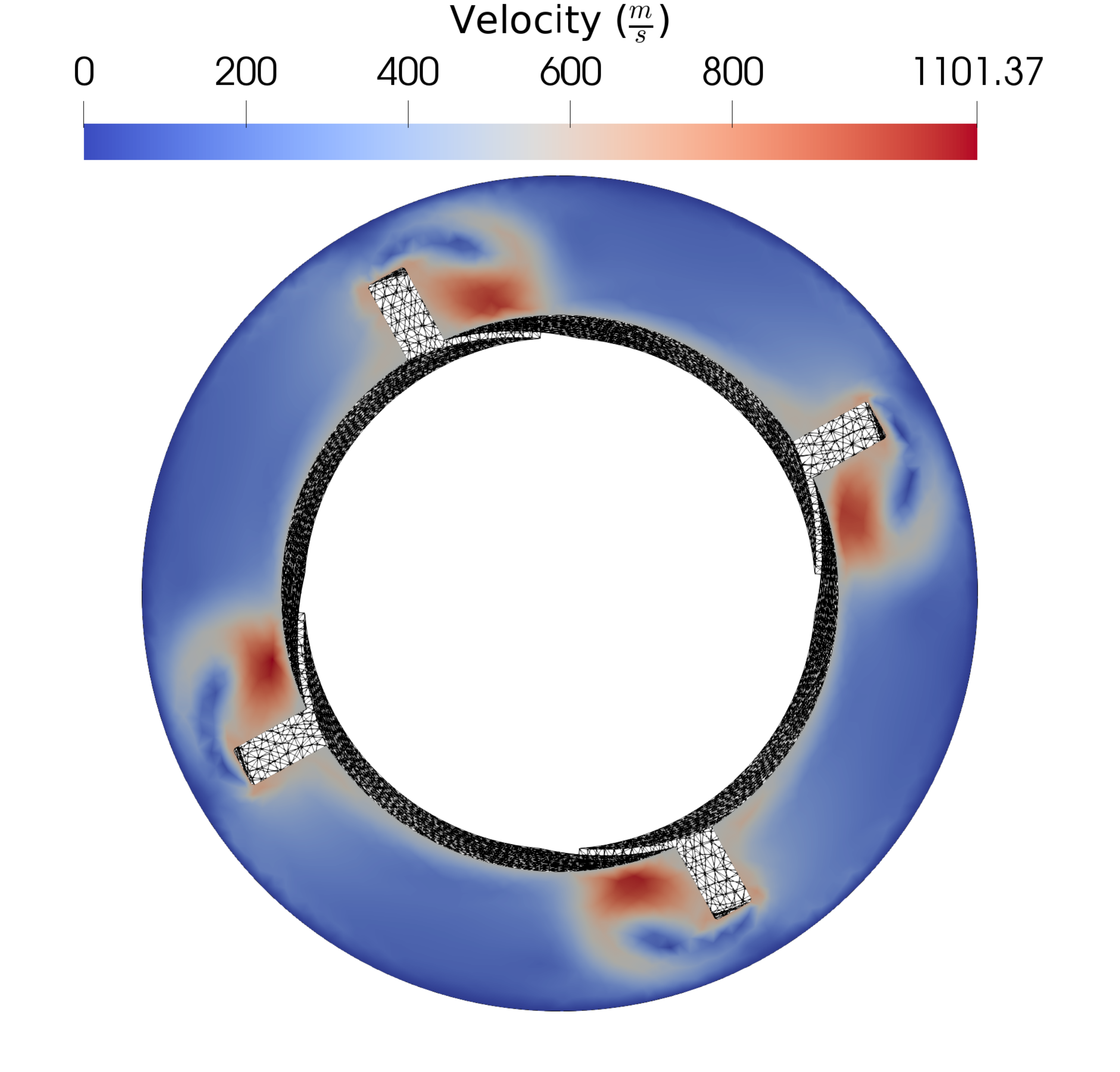}
		\captionsetup{justification=centering}
		\caption{Velocity (3d4n)}
		\label{Velocity3d4n}
	\end{subfigure}
	\hfill
	\begin{subfigure}{0.475\textwidth}
		\centering
		\includegraphics[width=1.0\linewidth]{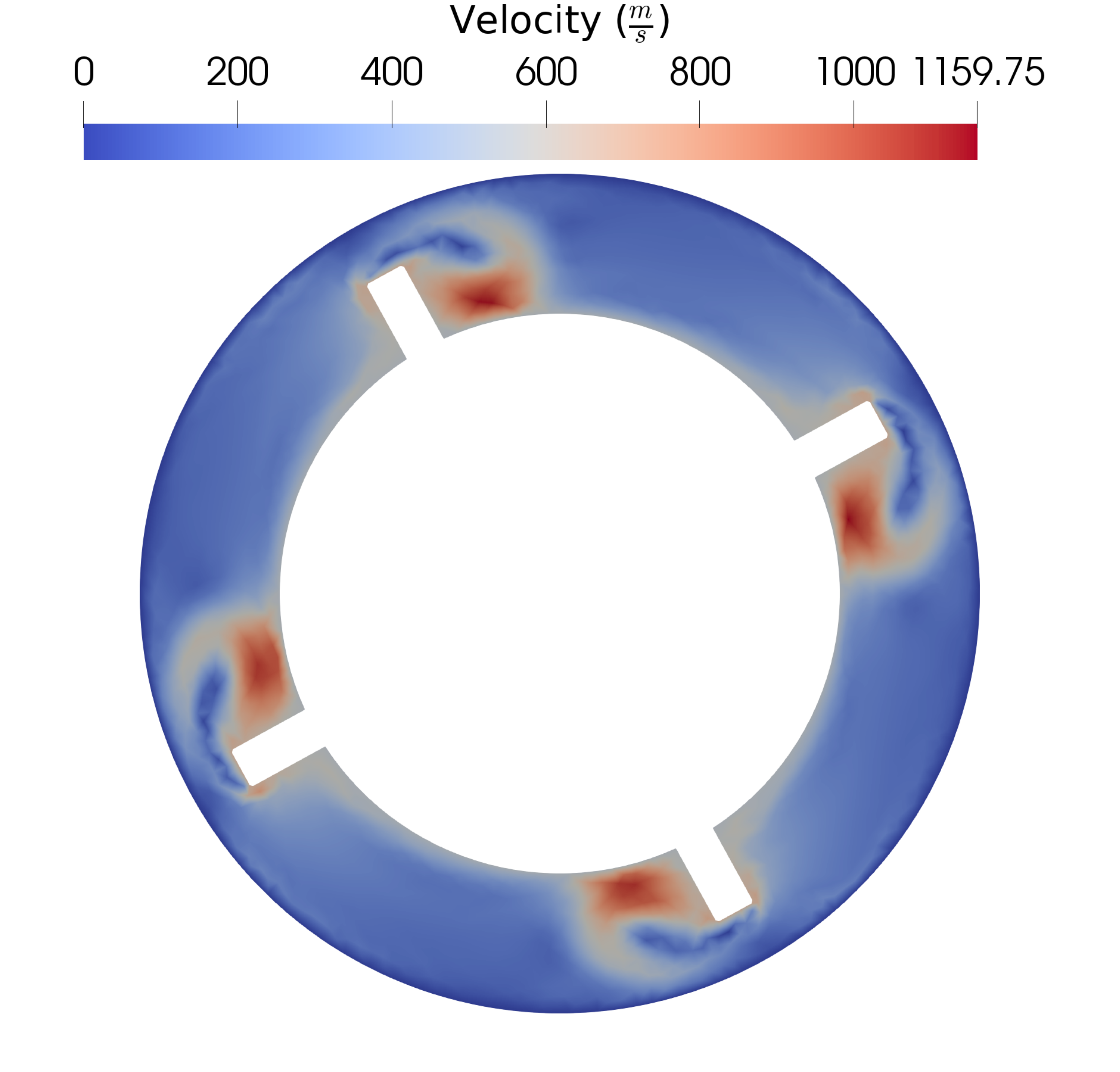}
		\captionsetup{justification=centering}
		\caption{Velocity (3d6n)}
		\label{Velocity3d6n}
	\end{subfigure}
	\\
	\begin{subfigure}{0.475\textwidth}
		\centering
		\includegraphics[width=1.0\linewidth]{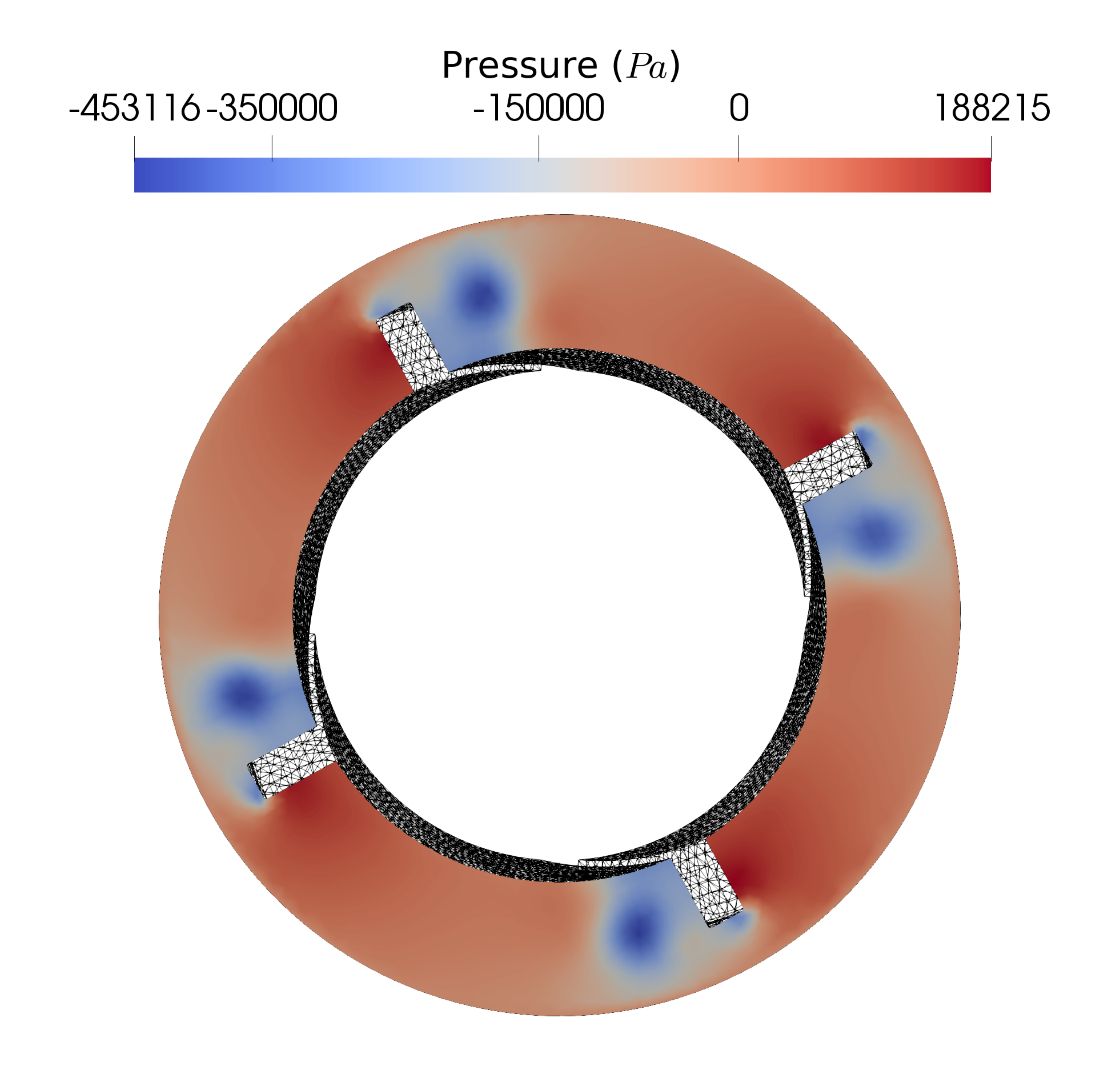}
		\captionsetup{justification=centering}
		\caption{Pressure (3d4n)}
		\label{Pressure3d4n}
	\end{subfigure}
	\hfill
	\begin{subfigure}{0.475\textwidth}
		\centering
		\includegraphics[width=1.0\linewidth]{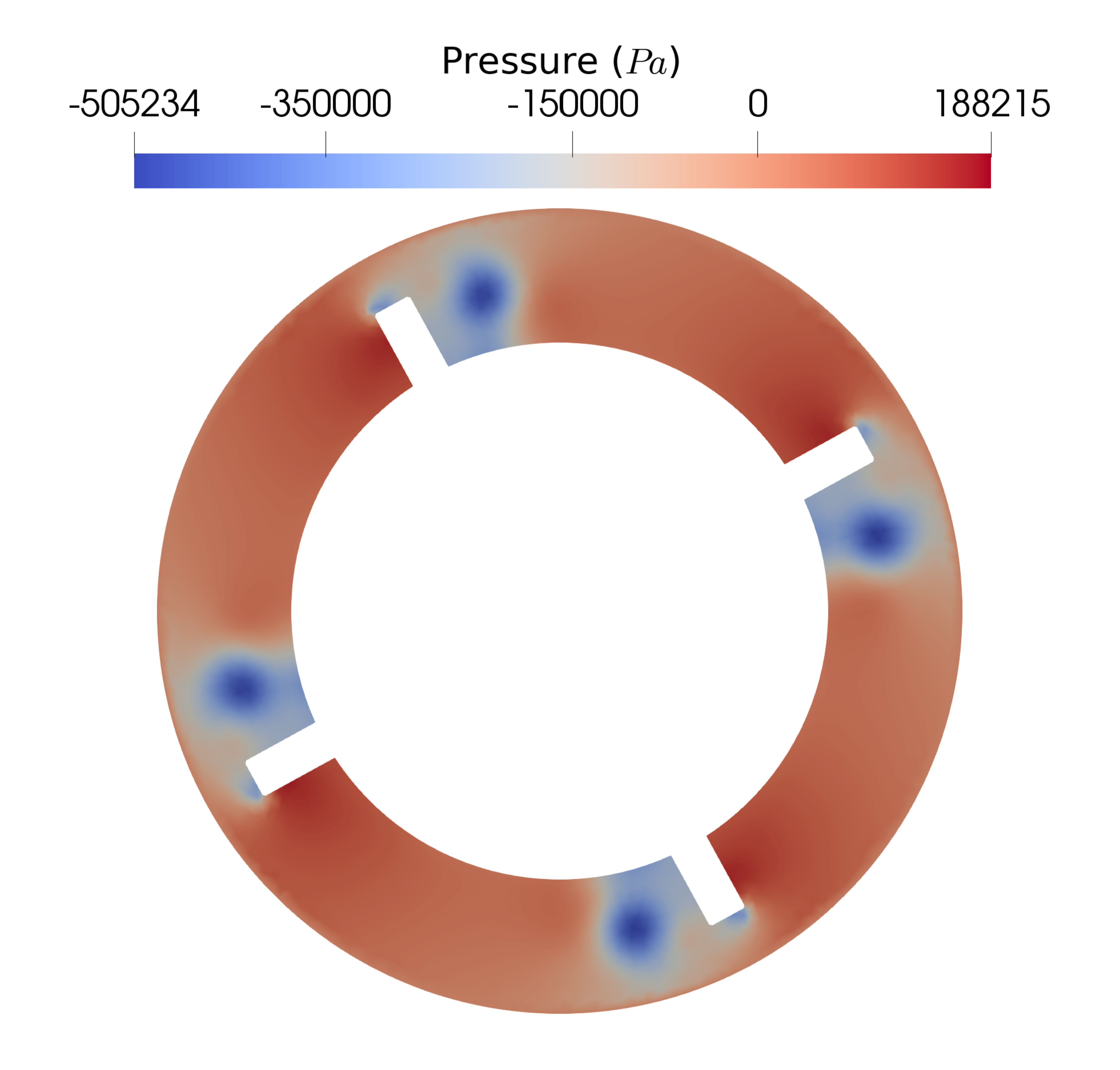}
		\captionsetup{justification=centering}
		\caption{Pressure (3d6n)}
		\label{Pressure3d6n}
	\end{subfigure}
	\caption{Comparison of the results obtained with a 3D space-time domain (\ref{Velocity3d4n} and \ref{Pressure3d4n}) and the concept of space-time slab (\ref{Velocity3d6n} and \ref{Pressure3d6n}).}
	\label{fig:Stirrer2D}
\end{figure}

Figure \ref{fig:Stirrer2D} shows the results of the computation with the UST discretization in comparison to the one using the time-slab concept, combined with the ALE formulation. As we can observe, vortices detach immediately at the tips of the blades in both simulations. The results of both computations qualitatively match. This promising comparison endorses the use of UST discretizations for problems with complex topology changes, i.e., including contact \cite{von2021four}, where usual numerical methods cannot be easily applied for predictions.

\subsection{3D Stirrer}
\label{3DStirrer}
In this section, we extend the previous benchmark case in three space dimensions. Hence, we investigate the three-dimensional flow in a short cylinder enclosing a rotating stirrer with thickness of \(\SI{0.1}{}\). The cross section of the stirrer is the same as in Figure \ref{fig:DimensionsStirrer2D}. The UST discretization consists of \(\SI{1673344}{}\) pentatopes, whereas the spatial grid for the ALE formulation contains \(\SI{24608}{}\) tetrahedra. The setup regarding the initial position of the stirrer, the boundary conditions, angular velocity, and material properties is the same as in Section \ref{2DStirrer}.

As before, the stirrer starts to rotate instantaneously in both simulations. In the simulation with the UST discretization, we just perform Newton-Raphson iterations to linearize the Navier-Stokes equations, and the system converges within \(\SI{19}{}\) iterations. In the simulation with the concept of space-time slab and ALE formulation, we use the same time-slab thickness as in the previous case. One time step usually converges within seven Newton-Raphson iterations.

\begin{figure}[!htb]
	\centering
	\begin{subfigure}{0.475\textwidth}
		\centering
		\includegraphics[width=1.0\linewidth]{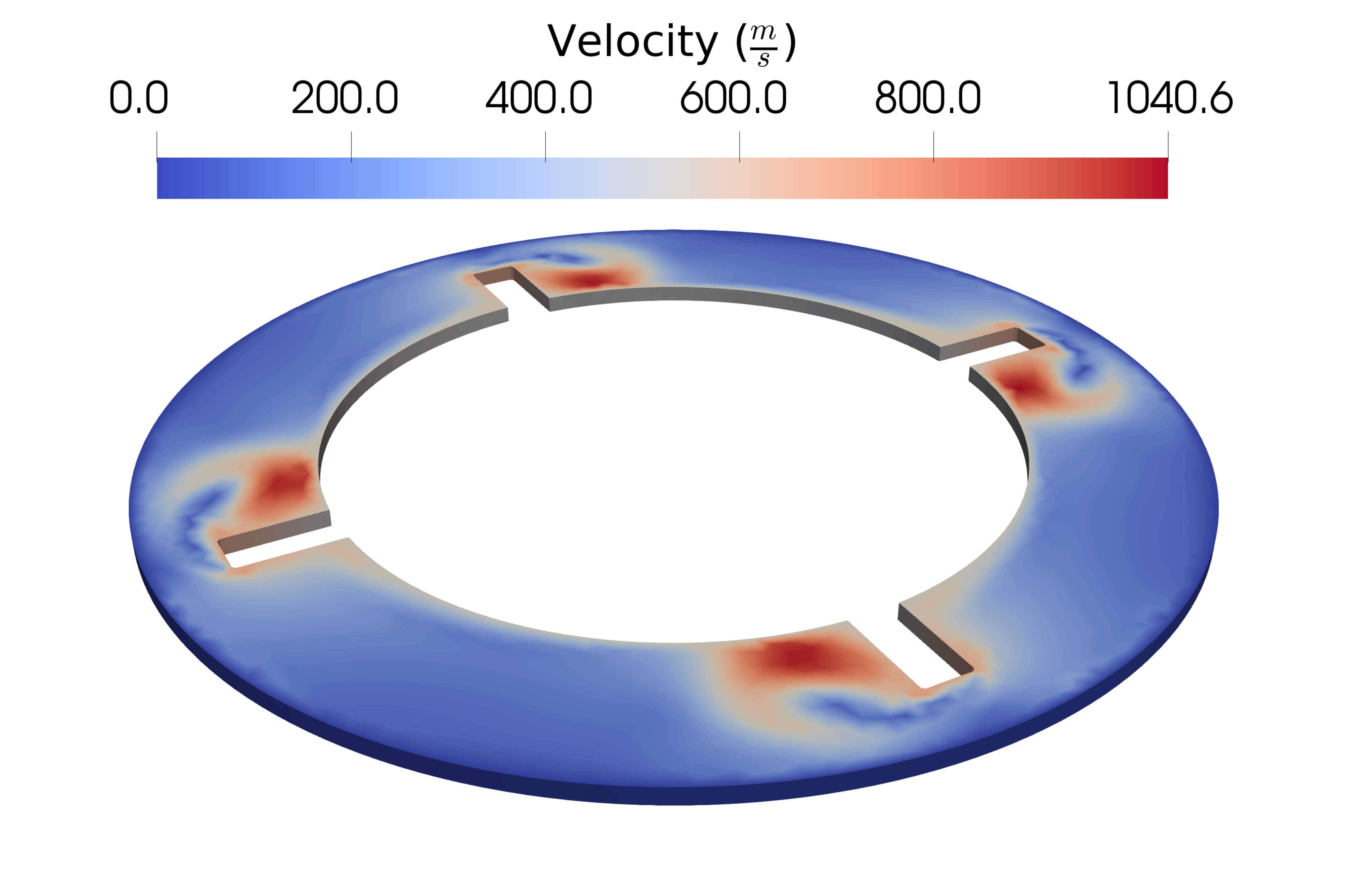}
		\captionsetup{justification=centering}
		\caption{Velocity (4d5n)}
		\label{Velocity4d5n}
	\end{subfigure}
	\hfill
	\begin{subfigure}{0.475\textwidth}
		\centering
		\includegraphics[width=1.0\linewidth]{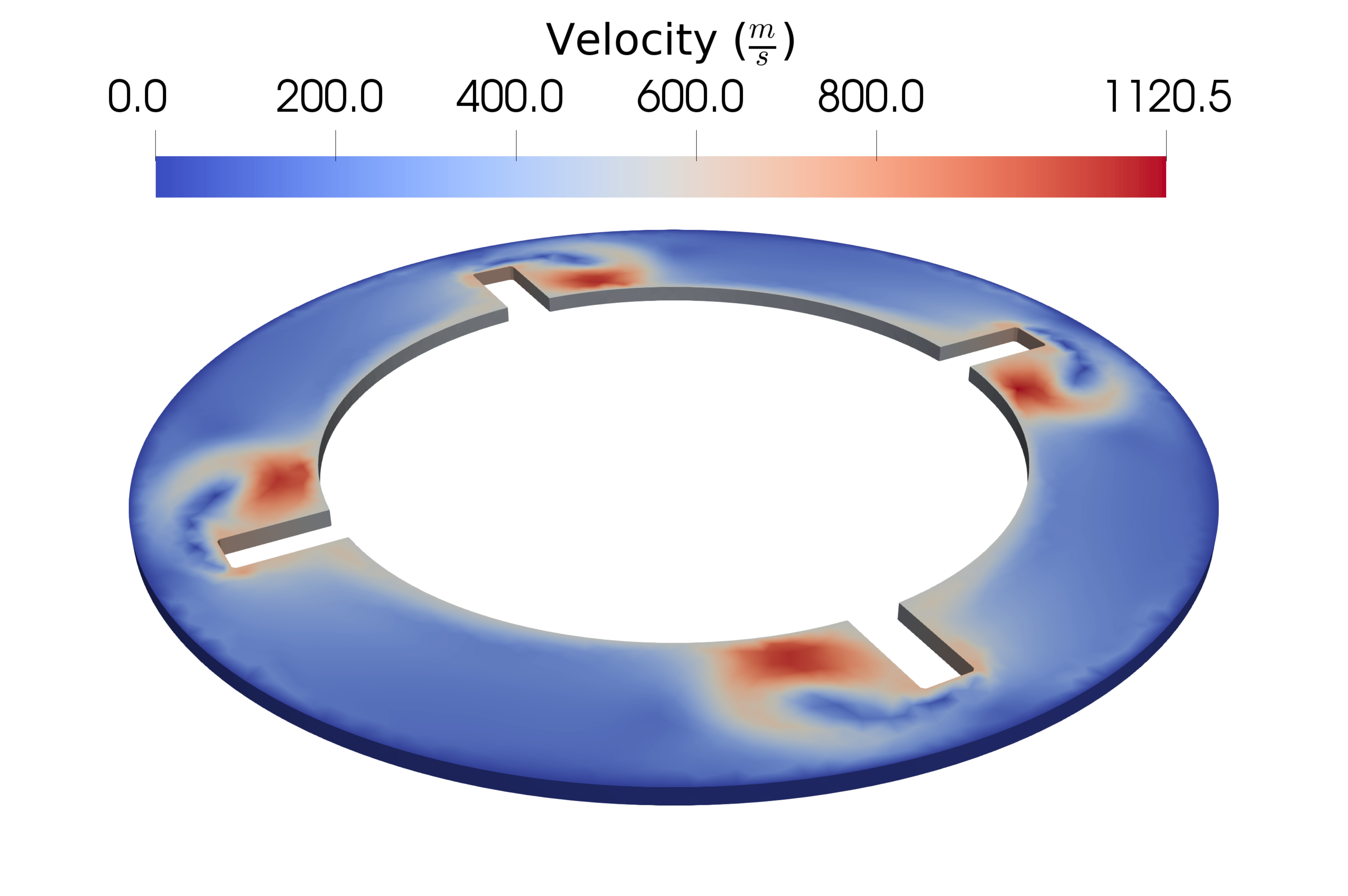}
		\captionsetup{justification=centering}
		\caption{Velocity (4d8n)}
		\label{Velocity4d8n}
	\end{subfigure}
	\\
	\begin{subfigure}{0.475\textwidth}
		\centering
		\includegraphics[width=1.0\linewidth]{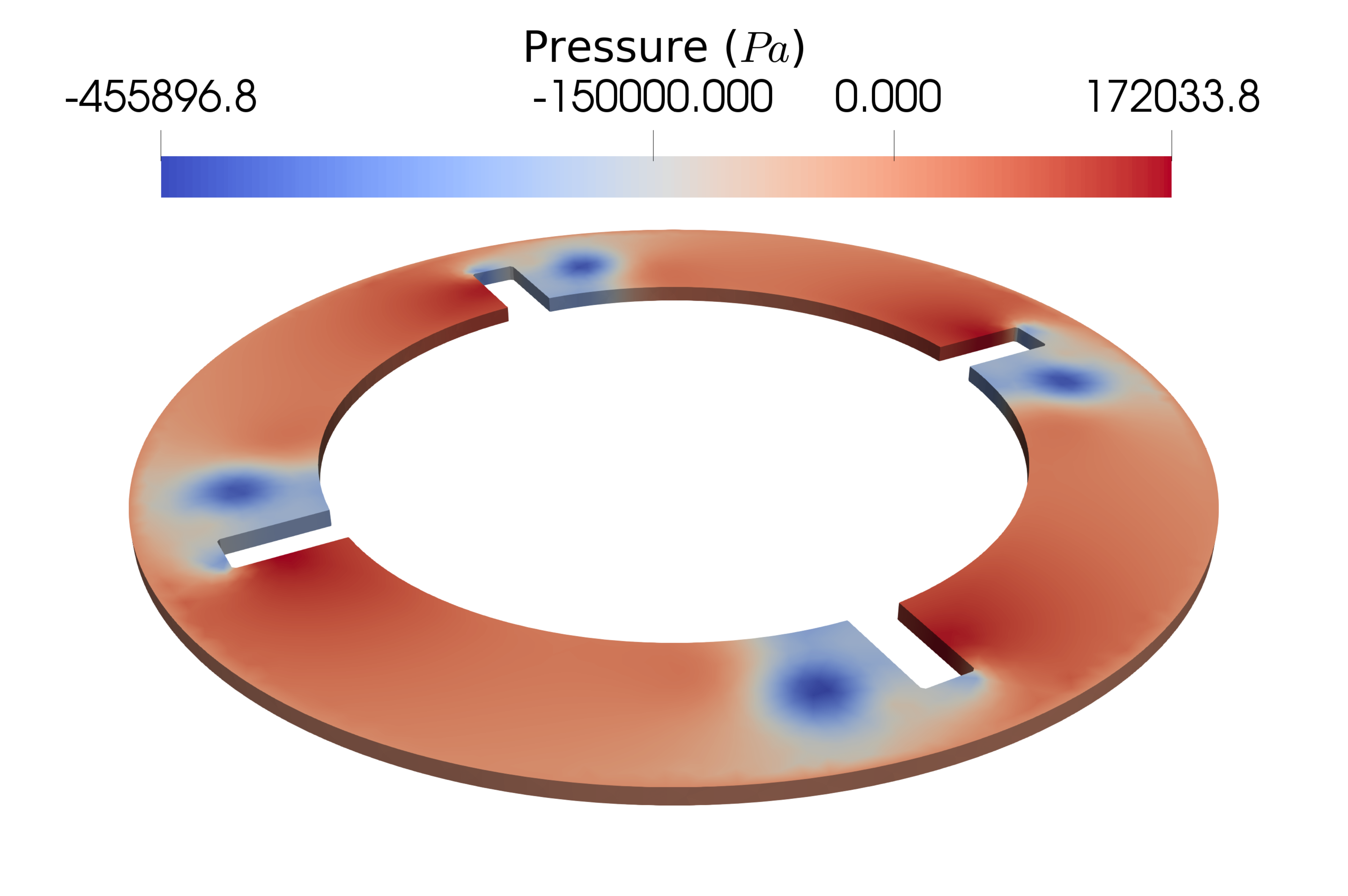}
		\captionsetup{justification=centering}
		\caption{Pressure (4d5n)}
		\label{Pressure4d5n}
	\end{subfigure}
	\hfill
	\begin{subfigure}{0.475\textwidth}
		\centering
		\includegraphics[width=1.0\linewidth]{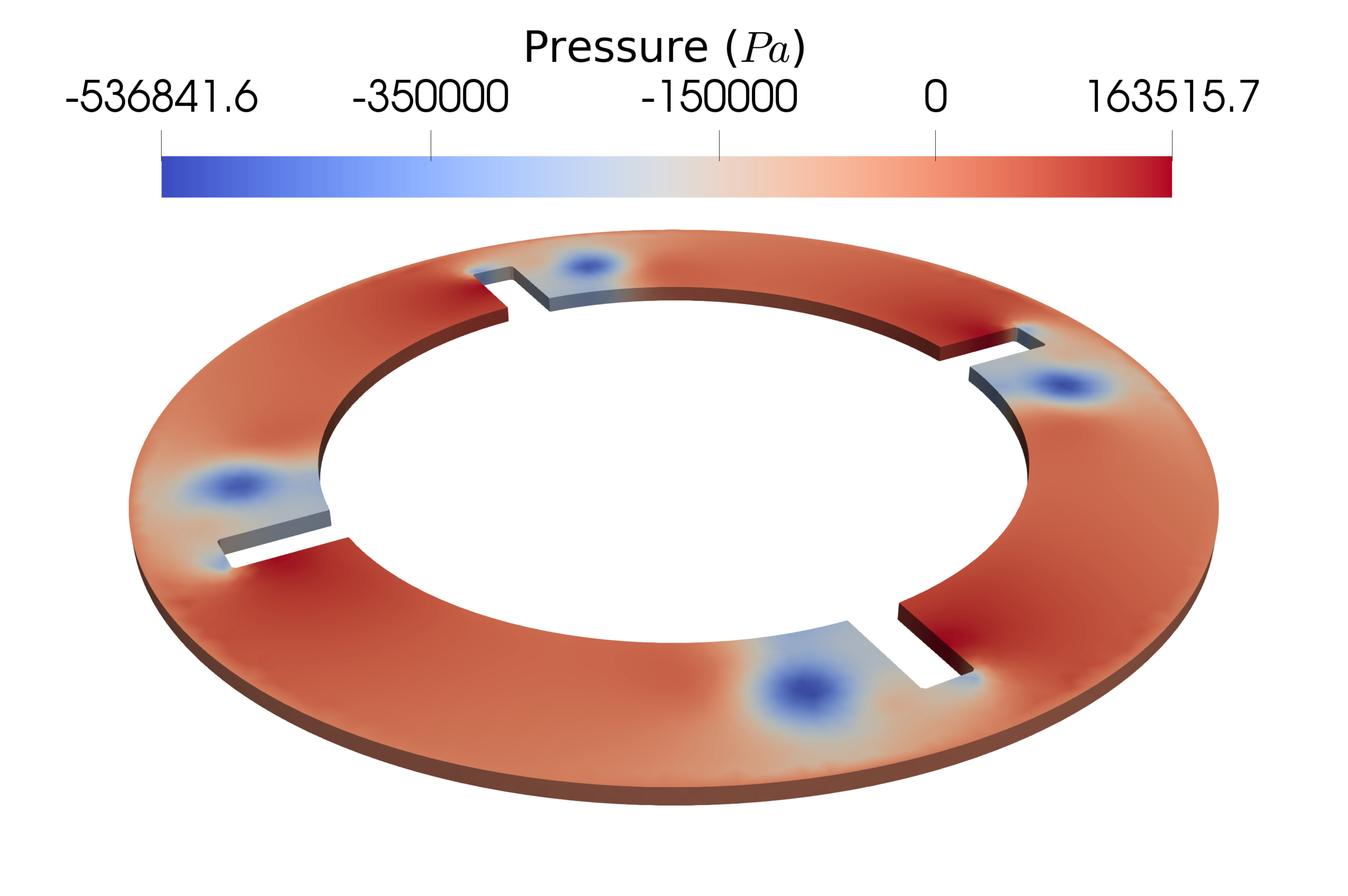}
		\captionsetup{justification=centering}
		\caption{Pressure (4d8n)}
		\label{Pressure4d8n}
	\end{subfigure}
	\caption{Comparison of the results obtained with a 4D space-time domain (\ref{Velocity4d5n} and \ref{Pressure4d5n}) and the concept of space-time slab (\ref{Velocity4d8n} and \ref{Pressure4d8n}).}
	\label{fig:Stirrer3D}
\end{figure}

Figure \ref{fig:Stirrer3D} shows the results on a slice of the space-time domain. In space-time approaches, it is not trivial to visualize the simulation results, especially in the cases of 4D SST or UST meshes. When the concept of a time slab in combination with FST meshes is used, we can simply show the results of the top or bottom of a space-time grid. However, when UST meshes are employed, we need to visualize intermediate slices of the space-time domain, especially in 4D cases, by extracting information from inside the domain. Therefore, a mesh projector algorithm is proposed by \citet{Fernandez2019} to solve this problem and to extract information within space-time domains. Further description can also be found in the work of \citet{Karyofylli:816042}. We compare the computation of the UST discretization with that using the time-slab concept combined with the ALE formulation. As we can see in Figure \ref{fig:Stirrer3D}, the velocity field and pressure distribution in the 3D case show the same pattern as in the 2D case. Furthermore, the results of the two approaches are in good agreement, strengthening our argument that UST discretizations provide reliable results, not only in 2D but also in 3D space, and can be used for problems with time-variant topology. 

\section{Conclusion}
We presented the computation of a rotating stirrer in two and three space dimensions. For the simulations, we used both a UST discretization that resolves the total movement of the stirrer and the concept of space-time slab in combination with mesh movement techniques. Comparing these two approaches is very promising and confirms that UST discretization can be used for cases with topology changes, while providing reliable predictions. Future work includes the improvement of our space-time mesh generator. A possible way can be the robust algorithm for triangulating submanifolds proposed by \citet{Boissonnat2020}.

\subsubsection*{Acknowledgement}
This work was mainly conducted during the Ph.D. studies of Violeta Karyofylli. Therefore, the authors gratefully acknowledge the sponsorship and support of the Deutsche Forschungsgemeinschaft e.V. (DFG, German Research Foundation) within the framework of the Collaborative Research Centre SFB1120-236616214 “Bauteilpräzision durch Beherrschung von Schmelze und Erstarrung in Produktionsprozessen”. The computations were conducted on computing clusters provided by the RWTH Aachen University IT Center and by the J\"ulich Aachen Research Alliance (JARA).

\bibliographystyle{unsrtnat}
\bibliography{referencesCorrectedStyle}  






\end{document}